# On the Capacity Region of the Cognitive Interference Channel with Unidirectional Destination Cooperation

Hsuan-Yi Chu and Hsuan-Jung Su
Graduate Institute of Communication Engineering
Department of Electrical Engineering
National Taiwan University, Taipei, Taiwan
Email: b95901210@ntu.edu.tw, hjsu@cc.ee.ntu.edu.tw

*Abstract*—The cognitive interference channel with unidirectional destination cooperation (CIFC-UDC) is a variant of the cognitive interference channel (CIFC) where the cognitive (secondary) destination not only decodes the information sent from its sending dual but also helps enhance the communication of the primary user. This channel is an extension of the original CIFC to achieve a win-win solution under the coexistence condition. The CIFC-UDC comprises a broadcast channel (BC), a relay channel (RC), as well as a partially cooperative relay broadcast channel (PCRBC), and can be degraded to any one of them. In this paper, we propose a new achievable rate region for the discrete memoryless CIFC-UDC which improves the previous results and includes the largest known rate regions of the BC, the RC, the PCRBC and the CIFC. A new outer bound is presented and proved to be tight for two classes of the CIFC-UDCs, resulting in the characterization of the capacity region.

## I. INTRODUCTION

The present regulation policy of spectrum utilization is to divide the spectrum into licensed lots to be allocated to different entities for exclusive use [13]. However, as the number of wireless devices has increased tremendously over the last few decades, the availability of wireless spectrum has become severely limited. This fact has led to a situation where new services have difficulty obtaining spectrum licenses and cannot be accommodated under the current regulation policy. This situation has been termed as "spectrum gridlock" [13] and considered as one of the main factors which may thwart further advancement and development of wireless technologies.

In recent years, various strategies for overcoming spectrum gridlock have been proposed [13]. Among them, cooperative communication has been envisioned to surmount this issue. Devices can cooperate to share spectrum, time slots and resources, which leads to more efficient communications. In the existing literature, the *cognitive interference channel* (*CIFC*), depicted in Fig. 2 of [12], has been one of the most intensively studied collaborative channel models. The CIFC is an asymmetrical two-user (primary user (PU) and cognitive user (CU)) communication network where the sender of the CU, with full knowledge of the message of the PU, and the sender of the PU attempt to communicate with their respective destinations through the common medium simultaneously. This channel model was first proposed and investigated by Devroye et al. [9] who presented the first achievable rate region for the *discrete memoryless CIFC* (*DM-CIFC*). Their work further demonstrated that by establishing the cooperation me-chanism between the two senders (termed as *genie-aided-cognition* originally) the interference caused at secondary destination can be effectively mitigated via *Gel'fand-Pinsker precoding* (*GP precoding*) [6]. As a result, the achievable rate region of the CIFC is significantly larger than that of the classical interference channel where, contrary to the cooperative communication in the CIFC, two senders work independently [7]. Since then, the CIFC has attracted a great deal of attentions, and several achievable rate regions as well as the capacity results for some special classes have been proved. A clear summary can be found in [16].

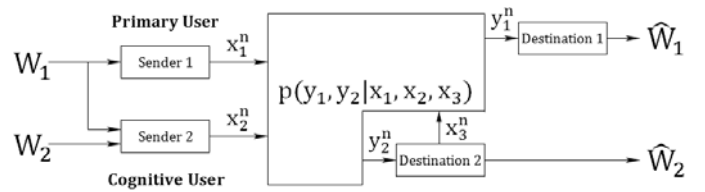

Figure 1. The channel model of the cognitive interference channel with unidirectional destination cooperation (CIFC-UDC).

In this work, we consider an extension of the original CIFC, the *cognitive interference channel with unidirectional destination cooperation* (*CIFC-UDC*), depicted in Figure 1. This channel model not only adopts the non-causal cognition mechanism in the original CIFC but also allows the secondary destination to participate in the communication between the sending and receiving sides of the PU. This channel model was proposed in [17][1] which derived the achievable rate region for the discrete memoryless case. The establishment of the asymmetrical relationship between the two destinations is motivated by the following practical communication scenario. In licensed bands, the CU is usually used to model the unlicensed users (secondary users). As a "spectrum borrower", the CU is not allowed to cause interference to the license user, i.e., the PU. In addition, if possible, the CU can try to provide some benefits for the PU. If this can be done, it is possible for both the CU and the PU to benefit and attain a win-win solution. In the CIFC-UDC, the benefit for the PU is evaluated through the capacity gain from the unidirectional destination cooperation[2] and was demonstrated in [17].

---

[1] In [17], the CIFC-UDC was called the *interference channel with degraded message sets with unidirectional destination cooperation* (*IC-DMS-UDC*).

[2] Destination (or receiver) cooperation in the classical interference channel has also been proved to bring capacity gain in [15] and [18]. The distinction

From an information-theoretic perspective, the CIFC-UDC is a "compound channel model" which includes several existing models: the *broadcast channel* (*BC*), the *relay channel* (*RC*), the *partially cooperative relay broadcast channel* (*PCRBC*) and the CIFC (e.g., by setting the output of destination 2 to null, the CIFC-UDC is degraded to the CIFC.). The main result of this paper is to derive a new achievable rate region for the discrete memoryless CIFC-UDC. This region can further be proved to subsume the previous result for the CIFC-UDC [17], the *Marton region* for the BC [4], the largest known rate for the RC, the *CMG rate* [8], and the largest known regions for the PCRBC [14] and the CIFC [16]. More importantly, a potentially larger achievable rate region for the PCRBC is also derived from our region[3].

The rest of this paper is structured as follows. In Section II, the notations and the channel model of the CIFC-UDC are defined. In Section III, we present our new achievable rate region for the CIFC-UDC. Additionally, the operation of the interference mitigation via the unidirectional destination cooperation is discussed. This interference mitigation strategy can be considered a complementary idea of the non-causal cognition proposed by Devroye et al. [9] that employs the GP precoding to alleviate the interference caused by the PU at the secondary destination. In Section IV, a new outer bound is presented, and the capacity results for two special classes of the CIFC-UDCs are derived. Lastly, in Section V, we conclude this paper.

## II. NOTATIONS, DEFINITIONS AND CHANNEL MODEL

### A. Notations

The following notational convention will be adopted throughout this paper. We use capital letters to denote random variables and lower case letters to denote their corresponding realizations. We adopt the notational convenience $p_{Y|X}(y|x) = p(y|x)$ to drop the subscript of the probability distribution. We use $x^n$ to represent the vector $(x_1, x_2, ..., x_n)$. Moreover, $X - Y - Z$ denotes that $X$, $Y$ and $Z$ form a Markov chain, and the notation $|\cdot|$ is used to denote the cardinality of a set. For the information theoretic quantities such as entropy, mutual information, etc., we follow the notations defined in [3].

### B. Definitions and Channel Model

The cognitive interference channel with unidirectional destination cooperation, depicted in Figure 1, is a channel model where sender 1 sends a message $w_1 \in W_1 = \{1,2,...,|W_1|\}$ to its destination with the help of destination 2 in n channel transmissions, and sender 2, with full knowledge of the message of sender 1, sends a message $w_2 \in W_2 = \{1,2,...,|W_2|\}$ to its destination in n channel transmissions. In addition, we focus on the *discrete memoryless CIFC-UDC* (*DM-CIFC-UDC*). The CIFC-UDC is said to be discrete memoryless in the sense that

$$p(y_1^n, y_2^n | x_1^n, x_2^n, x_3^n) = \prod_{t=1}^{n} p(y_{1_t}, y_{2_t} | x_{1_t}, x_{2_t}, x_{3_t}) \quad (II.1)$$

where the subscript t denotes the discrete time instant.

---

between these two works lies in the fact that the cooperative links in the former share the same band as the links in the interference channel, while those in the latter are orthogonal to each other as well as the links in the interference channel.

[3] Due to limited space, the detailed proofs of our results are not given in this paper. They can be found in [19].

*Definition 1:* The DM-CIFC-UDC is described by a tuple $(\mathcal{X}_1, \mathcal{X}_2, \mathcal{X}_3, \mathcal{Y}_1, \mathcal{Y}_2, p(y_1, y_2 | x_1, x_2, x_3))$, where $\mathcal{X}_1$, $\mathcal{X}_2$ and $\mathcal{X}_3$ denote the channel input alphabets, $\mathcal{Y}_1$ and $\mathcal{Y}_2$ denote the channel output alphabets, and $p(y_1, y_2 | x_1, x_2, x_3)$ denotes the transition probability. Moreover, $x_1$, $x_2$ and $x_3$ are channel inputs from sender 1, sender 2 and destination 2 respectively. $y_1$ and $y_2$ denote the channel outputs at destination 1 and destination 2 respectively.

Next we present the following definitions with regard to the existence of code, the achievable rate region and the capacity region for the DM-CIFC-UDC.

*Definition 2:* A $(|W_1|, |W_2|, n, P_e)$ code consists of:
- The message set $W_k = \{1,2,...,|W_k|\}$ where $|W_k| = 2^{nR_k}$, $k = 1,2$. It is assumed that the messages $W_1$ and $W_2$ are independent and uniformly distributed.
- An encoding function for sender 1 to map $w_1 \in W_1$ to a codeword $x_1^n$:
$$f_1: w_1 \to x_1^n \quad (II.2)$$
- An encoding function for sender 2 to map $w_1 \times w_2 \in W_1 \times W_2$ to a codeword $x_2^n$:
$$f_2: w_1 \times w_2 \to x_2^n \quad (II.3)$$
- A family of the encoding functions for destination 2 to map the preceding observations to the next transmitted symbol $x_{3_i}$:
$$f_{3_i}: (y_{2_1}, y_{2_2}, y_{2_3}, ..., y_{2_{i-1}}) \to x_{3_i} \text{ for } 1 \leq i \leq n \quad (II.4)$$
- The decoding function for decoder k to map $y_k^n$ to $\hat{w}_k \in W_k$.
$$g_k: y_k^n \to \hat{w}_k, k = 1,2 \quad (II.5)$$
- The average probability of error:
$$P_e \equiv \max\{P_{e_1}, P_{e_2}\} \quad (II.6)$$

where $P_{e_k}$ denotes the average probability of error of decoder k, k = 1,2. Furthermore, because it is assumed that the message pair $(w_1, w_2) \in (W_1, W_2)$ is equiprobable, $P_{e_k}$ can be computed as

$$P_{e_k} = \frac{1}{2^{n(R_1+R_2)}} \sum_{(w_1, w_2)} \Pr\{\hat{w}_k \neq w_k | (w_1, w_2) \text{ sent.}\} \quad (II.7)$$

*Definition 3:* A nonnegative rate pair $(R_1, R_2)$ is said to be achievable for the CIFC-UDC if there exists a $(|W_1|, |W_2|, n, P_e)$ code with $R_1 \leq \frac{1}{n} \log |W_1|$ and $R_2 \leq \frac{1}{n} \log |W_2|$ such that $P_e \to 0$ as $n \to \infty$.

*Definition 4:* The capacity region for the CIFC-UDC, denoted as $\mathcal{C}$, is the closure of the region of all the achievable rate pairs $(R_1, R_2)$. An achievable rate region, denoted as $\mathcal{R}$, is a subset of the capacity region.

## III. A NEW ACHIEVABLE RATE REGION FOR THE DM-CIFC-UDC

Since the DM-CIFC-UDC is a compound channel model with a variety of aspects, we apply a collection of coding strategies to derive our result, including: rate splitting, Gel'fand-Pinsker binning/precoding, Marton binning, generalized relaying strategy of Cover and El Gamal, interference forwarding, and the improved simback decoding. A brief summary of these coding strategies can be found in [19].

In the following, we present our new unified achievable rate region for the DM-CIFC-UDC.

*Theorem 1:* Let $\mathcal{P}$ denote the set of all joint probability distributions

$$p(u_{1p}, u_1, v_1, u_{2p}, u_2, v_{12}, v_2, x_1, x_2, x_3, y_1, y_2, \hat{y}_2) \quad \text{(III.1.A)}$$

which can be factored in the following form:

$$\begin{aligned}&p(u_{1p})p(u_1|u_{1p})p(v_1|u_{1p},u_1)p(u_{2p}|u_{1p})\\&\times p(u_2,v_{12},v_2|u_{1p},u_1,v_1,u_{2p})p(x_1|u_{1p},u_1,v_1)\\&\times p(x_2|u_{1p},u_1,v_1,u_{2p},u_2,v_{12},v_2)p(x_3|u_{1p},u_{2p})\\&\times p(y_1,y_2|x_1,x_2,x_3)p(\hat{y}_2|u_{1p},u_1,u_{2p},u_2,x_3,y_2)\end{aligned} \quad \text{(III.1.B)}$$

Let $\mathcal{R}(p)$ be the set of all nonnegative rate pairs $(R_1, R_2) = (R_{11} + R_{1P} + R_{1B}, R_{22} + R_{2P})$ such that the following constraints hold:

$$\begin{aligned}R_{11} \geq 0, R_{1P} \geq 0, R_{1B} \geq 0,\\ R_{22} \geq 0, R_{2P} \geq 0\end{aligned} \quad \text{(III.1.C)}$$

$$\begin{aligned}R'_{2P} &\geq A\\ R'_{1B} &\geq 0\\ R'_{22} &\geq I(V_1; V_2|U_{1p}, U_1, U_{2p}, U_2)\\ R'_{1B} + R'_{22} &\geq I(V_1, V_{12}; V_2|U_{1p}, U_1, U_{2p}, U_2)\end{aligned} \quad \text{(III.1.D)}$$

$$\begin{aligned}R_{1p} + R_{11} + L_{2P} + L_{1B} &\leq A + B + D - C\\ R_{11} + L_{2P} + L_{1B} &\leq A + B + E - C\\ L_{2P} + L_{1B} &\leq A + B + H - C\\ R_{11} + L_{1B} &\leq \min\begin{pmatrix}A+B+F-C,\\ A+G\end{pmatrix}\\ L_{1B} &\leq \min(J, B+I-C)\end{aligned} \quad \text{(III.1.E)}$$

$$\begin{aligned}R_{1P} + L_{2P} + L_{22} &\leq K\\ L_{2P} + L_{22} &\leq L\\ L_{22} &\leq M\end{aligned} \quad \text{(III.1.F)}$$

$$C \leq I(Y_1; X_3|U_{1p}, U_1, V_1, U_{2p}, U_2, V_{12}) + B \quad \text{(III.1.G)}$$

where

$L_{1B} = R_{1B} + R'_{1B}, \quad L_{2P} = R_{2P} + R'_{2P}, \quad L_{22} = R_{22} + R'_{22}$

$A = I(V_1; U_2|U_{1p}, U_1, U_{2p})$
$B = I(Y_1, V_1, V_{12}; \hat{Y}_2|U_{1p}, U_1, U_{2p}, U_2, X_3)$
$C = I(Y_2; \hat{Y}_2|U_{1p}, U_1, U_{2p}, U_2, X_3)$
$D = I(Y_1; U_{1p}, U_1, V_1, U_{2p}, U_2, V_{12}, X_3)$
$E = I(Y_1; V_1, U_{2p}, U_2, V_{12}, X_3|U_{1p}, U_1)$
$F = I(Y_1; V_1, V_{12}, X_3|U_{1p}, U_1, U_{2p}, U_2)$
$G = I(Y_1, \hat{Y}_2; V_1, V_{12}|U_{1p}, U_1, U_{2p}, U_2, X_3)$
$H = I(Y_1; U_{2p}, U_2, V_{12}, X_3|U_{1p}, U_1, V_1)$
$I = I(Y_1; V_{12}, X_3|U_{1p}, U_1, V_1, U_{2p}, U_2)$
$J = I(Y_1, \hat{Y}_2; V_{12}|U_{1p}, U_1, V_1, U_{2p}, U_2, X_3)$
$K = I(Y_2; U_1, U_2, V_2|U_{1p}, U_{2p}, X_3)$
$L = I(Y_2; U_2, V_2|U_{1p}, U_1, U_{2p}, X_3)$
$M = I(Y_2; V_2|U_{1p}, U_1, U_{2p}, U_2, X_3)$

Since destination 1 (2) is not interested in $W_{2P}$ ($W_{1P}$), some rate constraints can be dropped:

- The first constraint in (III.1.E) can be dropped if $R_{1p} = R_{11} = L_{1B} = 0$.
- The second constraint in (III.1.E) can be dropped if $R_{11} = L_{1B} = 0$.
- The third constraint in (III.1.E) can be dropped if $L_{1B} = 0$.
- The first constraint in (III.1.F) can be dropped if $L_{2P} = L_{22} = 0$.

Then the region $\mathcal{R} = \cup_{p(\cdot) \in \mathcal{P}} \mathcal{R}(p)$ is an achievable rate region for the DM-CIFC-UDC.

*Proof:* See [19]. ∎

*Remark 1:* Note that the achievable rate region defined by $\mathcal{R}$ is convex. Henceforth, no convex hull operation or time-sharing is necessary (see [19]).

*Remark 2:* As shown in [19], Theorem 1 subsumes the previous result for the CIFC-UDC [17], the *Marton region* for the BC [4], the largest known rate for the RC, *CMG rate* [8], and the largest known regions for the PCRBC [14] and the CIFC [16]. More importantly, a potentially larger achievable rate region for the PCRBC [19] can also be derived from Theorem 1.

*Remark 3:* While the unidirectional destination cooperation strategy presented in [17] only allows destination 2 to forward the public sub-message of the PU, $W_{1P}$, in this work, destination 2 is also able to forward the public sub-message of the CU, $W_{2P}$. This additional cooperation modality is actually an application of the *interference forwarding technique* [11] and plays an important role in the communication setting considered in Fig. 2 of [19]. (We only provide a brief explanation for this setting below. For more detailed discussion, we refer the readers to [19].)

As can be observed in Fig. 2 of [19], by forwarding the public sub-message of the CU, the reception of $W_2$ at destination 1 is enhanced, which helps destination 1 understand $W_2$ and subtract the interference associated with $W_2$ (i.e., to subtract $\sqrt{c_{21}} \times x_2^n(w_2)$ from $y_1^n$). More interestingly, by combing the idea above with the Gel'fand-Pinsker precoding, the interferences caused at both the destinations can be alleviated, which is impossible to be attained by the strategy of [17].

## IV. A NEW OUTER BOUND AND TWO CAPACITY RESULTS FOR THE DM-CIFC-UDC

In this section, we present a new outer bound for the DM-CIFC-UDC. Although this outer bound is not tight in general, for two special classes of the DM-CIFC-UDC, this outer bound can be proved to be achievable, resulting in the characterization of the capacity region.

*Theorem 2:* If $(R_1, R_2)$ lies in the capacity region for the DM-CIFC-UDC,

$$\begin{aligned}R_1 &\leq I(Y_1; X_1, X_2, X_3) & \text{(IV.1.A)}\\ R_1 &\leq I(Y_1; X_1, V_{12}, X_3) & \text{(IV.1.B)}\end{aligned}$$

$$\begin{aligned}R_2 &\leq I(Y_2; X_2|X_1, X_3) & \text{(IV.1.C)}\\ R_1 + R_2 &\leq I(Y_1, Y_2; X_1, X_2|X_3) & \text{(IV.1.D)}\\ R_1 + R_2 &\leq I(X_2; Y_2|X_1, V_{12}, X_3)\\ &\quad + I(X_1, V_{12}, X_3; Y_1) & \text{(IV.1.E)}\end{aligned}$$

taken over the union of all joint distributions $p(x_1, v_{12}, x_2, x_3, y_1, y_2)$.

*Proof:* In the Appendix. ∎

The capacity for the DM-CIFC-UDC has been an open problem since its inception. Here we focus on two special classes of the DM-CIFC-UDC, the *degraded DM-Z-CIFC-UDC* and the *semi-deterministic CIFC-UDC* in the *high-interference-gain regime*.

*Definition 5:* A DM-CIFC-UDC is said to be the DM-Z-CIFC-UDC if the transition probability $p(y_1, y_2 | x_1, x_2, x_3)$ can be factored as $p(y_1 | x_1, x_3) \times p(y_2 | x_1, x_2, x_3)$.

*Definition 6:* A DM-CIFC-UDC is said to be degraded if $(X_1, X_2) - (Y_2, X_3) - Y_1$ form a Markov chain.

*Theorem 3:* The capacity region for the degraded DM-Z-CIFC-UDC is the set of all nonnegative rate pairs $(R_1, R_2)$ satisfying the following constraints:

$$\begin{align} R_1 &\leq I(Y_1; X_1, X_3) & (\text{IV.2.A}) \\ R_2 &\leq I(Y_2; X_2 | X_1, X_3) & (\text{IV.2.B}) \\ R_1 + R_2 &\leq I(Y_2; X_1, X_2 | X_3) & (\text{IV.2.C}) \end{align}$$

taken over the union of all joint distributions $p(x_1, x_2, x_3, y_1, y_2)$.

*Proof:* See [19]. ∎

*Definition 7:* A DM-CIFC-UDC is said to be semi-deterministic if the transition probability distribution $p(y_2 | x_1, x_2, x_3)$ takes only the value 0 or 1.

*Definition 8:* A DM-CIFC-UDC is said to be in the high-interference-gain regime if

$$\begin{align} I(Y_2; X_1 | X_3) &\geq I(Y_1; X_1, X_3) & (\text{IV.3.A}) \\ I(Y_1; V_{12} | X_1, X_3) &\geq I(Y_2; V_{12} | X_1, X_3) & (\text{IV.3.B}) \end{align}$$

for all distributions $p(x_1, v_{12}, x_2, x_3, y_1, y_2)$.

*Theorem 4:* The capacity region for the semi-deterministic CIFC-UDC in the high-interference-gain regime is the set of all nonnegative rate pairs $(R_1, R_2)$ satisfying the following constraints:

$$\begin{align} R_1 &\leq I(Y_1; X_1, V_{12}, X_3) & (\text{IV.4.A}) \\ R_2 &\leq H(Y_2 | X_1, X_3) & (\text{IV.4.B}) \\ R_1 + R_2 &\leq I(Y_1; X_1, V_{12}, X_3) + H(Y_2 | X_1, V_{12}, X_3) & (\text{IV.4.C}) \end{align}$$

taken over the union of all joint distributions $p(x_1, v_{12}, x_2, x_3, y_1, y_2)$.

*Proof:* The converse part can be proved by Theorem 2 and using the fact that $H(Y_2 | X_1, X_2, X_3) = 0$.

For the achievability part, by setting $v_1$, $u_{2p}$, $u_2$ and $\hat{y}_2$ as $\emptyset$, $u_1$ as $x_1$ and $u_{1p}$ as $x_3$ and performing *Fourier-Motzkin elimination* [1], the set of all nonnegative rate pairs $(R_1, R_2)$ satisfying the following constraints is an achievable rate region for the DM-CIFC-UDC:

$$\begin{align} R_1 &\leq I(Y_1; X_1, V_{12}, X_3) & (\text{IV.5.A}) \\ R_2 &\leq I(Y_2; V_2 | X_1, X_3) & (\text{IV.5.B}) \\ R_2 &\leq I(Y_2; V_2 | X_1, X_3) + I(Y_1; V_{12} | X_1, X_3) \\ &\quad - I(V_{12}; V_2 | X_1, X_3) & (\text{IV.5.C}) \\ R_1 + R_2 &\leq I(Y_1; V_{12} | X_1, X_3) + I(Y_2; X_1, V_2 | X_3) \\ &\quad - I(V_{12}; V_2 | X_1, X_3) & (\text{IV.5.D}) \\ R_1 + R_2 &\leq I(Y_1; X_1, V_{12}, X_3) + I(Y_2; V_2 | X_1, X_3) \\ &\quad - I(V_{12}; V_2 | X_1, X_3) & (\text{IV.5.E}) \end{align}$$

taken over the union of all joint distributions $p(x_1, v_{12}, v_2, x_2, x_3, y_1, y_2)$

In addition, since $Y_2$ is a deterministic function of the inputs, and the sender of the CU, with the knowledge of $W_1$ and $W_2$, knows these inputs, we can choose $V_2 = Y_2$. Thus, (IV.5) can be rewritten as:

$$R_1 \leq I(Y_1; X_1, V_{12}, X_3) \quad (\text{IV.6.A})$$

$$\begin{align} R_2 &\leq H(Y_2 | X_1, X_3) & (\text{IV.6.B}) \\ R_2 &\leq H(Y_2 | X_1, X_3) + I(Y_1; V_{12} | X_1, X_3) \\ &\quad - I(Y_2; V_{12} | X_1, X_3) & (\text{IV.6.C}) \\ R_1 + R_2 &\leq I(Y_1; V_{12} | X_1, X_3) + H(Y_2 | X_3) \\ &\quad - I(Y_2; V_{12} | X_1, X_3) & (\text{IV.6.D}) \\ R_1 + R_2 &\leq I(Y_1; X_1, V_{12}, X_3) + H(Y_2 | X_1, X_3) \\ &\quad - I(Y_2; V_{12} | X_1, X_3) & (\text{IV.6.E}) \end{align}$$

Furthermore, it can be observed that if (IV.3.A) and (IV.3.B) hold, (IV.6.C) and (IV.6.D) can be dropped. At last, after some manipulations, the achievability can be proved, and the capacity region for this class is derived. ∎

## V. CONCLUSION

In this paper, a new achievable rate region for the DM-CIFC-UDC is derived. This region can further be proved to include the previous result for the CIFC-UDC, the largest known achievable rate / regions for the broadcast channel, the relay channel, the partially cooperative relay broadcast channel and the cognitive interference channel. More importantly, a potentially larger achievable rate region for the partially cooperative relay broadcast channel can also be derived from our theorem.

In addition, how to assist the decoder of the PU to alleviate the interference caused by the transmission of the CU via the unidirectional destination cooperation is also investigated. This interference mitigation strategy can be considered as a complementary idea of the non-causal cognition proposed by Devroye *et al.* that employs Gel'fand-Pinsker precoding to alleviate the interference caused by the PU at the secondary destination.

A new outer bound of the capacity region for the CIFC-UDC is also derived in this paper. Although this outer bound is not tight in general, for two classes of the DM-CIFC-UDC, our achievable rate region meets this outer bound, resulting in the characterization of the capacity region.

## APPENDIX
## PROOF OF THEOREM 2

Consider a code $(|W_1|, |W_2|, n, P_e)$ with average probability of error $P_e \to 0$. The probability distribution on the joint ensemble space $W_1 \times W_2 \times \mathcal{X}_1^n \times \mathcal{X}_2^n \times \mathcal{X}_3^n \times \mathcal{Y}_1^n \times \mathcal{Y}_2^n$ is given by

$$\begin{align} &p(w_1, w_2, x_1^n, x_2^n, x_3^n, y_1^n, y_2^n) \\ &= p(w_1)\, p(w_2)\, p(x_1^n | w_1) p(x_2^n | w_1, w_2) \\ &\quad \times \prod_{i=1}^{n} \left( I\left(x_{3_i} = f_{3_i}(y_2^{i-1})\right) \times p(y_{1_i}, y_{2_i} | x_{1_i}, x_{2_i}, x_{3_i}) \right) \quad (A1) \end{align}$$

where $I(\cdot)$ is the indicator function which equals to 1 if its argument is true and equals to 0 otherwise. Additionally, $p(x_1^n | w_1)$ and $p(x_2^n | w_1, w_2)$ are either 0 or 1.

According to Fano's inequality [2], we have

$$\begin{align} H(W_1 | Y_1^n) &\leq n R_1 P_e + 1 \triangleq n \varepsilon_1 & (A2) \\ H(W_2 | Y_2^n) &\leq n R_2 P_e + 1 \triangleq n \varepsilon_2 & (A3) \end{align}$$

where $\varepsilon_1$ and $\varepsilon_2$ goes to zero if $P_e \to 0$. We can bound the rate $R_1$ as

$R_1 = H(W_1)$
$= I(W_1; Y_1^n) + H(W_1 | Y_1^n)$

$$\leq \sum_{i=1}^{n} I(Y_{1_i}; W_1 | Y_{1_{i+1}}^n) + n\varepsilon_1$$

$$= \left( \sum_{i=1}^{n} H(Y_{1_i} | Y_{1_{i+1}}^n) - H(Y_{1_i} | Y_{1_{i+1}}^n, W_1) \right) + n\varepsilon_1$$

$$\leq \left( \sum_{i=1}^{n} H(Y_{1_i}) - H(Y_{1_i} | Y_{1_{i+1}}^n, Y_2^{i-1}, X_{1_i}, X_{3_i}, W_1) \right) + n\varepsilon_1 \quad (A4)$$

where (A4) follows since conditioning does not increase entropy. Besides, we define the following auxiliary random variable $V_{12_i} \triangleq (Y_{1_{i+1}}^n, Y_2^{i-1}, W_1)$. Therefore, (A4) can be rewritten as:

$$\left( \sum_{i=1}^{n} H(Y_{1_i}) - H(Y_{1_i} | X_{1_i}, V_{12_i}, X_{3_i}) \right) + n\varepsilon_1$$

$$= \left( \sum_{i=1}^{n} I(Y_{1_i}; X_{1_i}, V_{12_i}, X_{3_i}) \right) + n\varepsilon_1 \quad (A5)$$

Further we bound the sum rate $R_1 + R_2$.

$$R_1 + R_2 = H(W_1) + H(W_2)$$
$$= I(W_1; Y_1^n) + H(W_1 | Y_1^n) + I(W_2; Y_2^n) + H(W_2 | Y_2^n)$$
$$\leq I(Y_1^n; W_1) + I(W_2; Y_2^n, W_1) + n(\varepsilon_1 + \varepsilon_2)$$
$$= I(Y_1^n; W_1) + I(W_2; Y_2^n | W_1) + n\varepsilon \quad (A6)$$

where (A6) follows from the independence between $W_1$ and $W_2$ and from $\varepsilon \triangleq \varepsilon_1 + \varepsilon_2$. Moreover,

$$I(Y_1^n; W_1) + I(W_2; Y_2^n | W_1) + n\varepsilon$$
$$= \sum_{i=1}^{n} I(Y_{1_i}^n; W_1, Y_2^{i-1}) - I(Y_{1_{i+1}}^n; W_1, Y_2^i)$$
$$+ I(Y_{2_i}; W_2 | W_1, Y_2^{i-1}) + n\varepsilon \quad (A7)$$

$$= \sum_{i=1}^{n} \left( I(Y_{1_{i+1}}^n; W_1, Y_2^{i-1}) + I(Y_{1_i}; W_1, Y_2^{i-1} | Y_{1_{i+1}}^n) \right)$$
$$- \left( I(Y_{1_{i+1}}^n; W_1, Y_2^{i-1}) + I(Y_{1_{i+1}}^n; Y_{2_i} | W_1, Y_2^{i-1}) \right)$$
$$+ I(Y_{2_i}; W_2 | W_1, Y_2^{i-1}) + n\varepsilon \quad (A8)$$

$$= \sum_{i=1}^{n} I(Y_{1_i}; W_1, Y_2^{i-1} | Y_{1_{i+1}}^n) - I(Y_{1_{i+1}}^n; Y_{2_i} | W_1, Y_2^{i-1})$$
$$+ I(Y_{2_i}; W_2 | W_1, Y_2^{i-1}) + n\varepsilon$$

$$= \sum_{i=1}^{n} I(Y_{1_i}; W_1, Y_2^{i-1} | Y_{1_{i+1}}^n) - I(W_2, Y_{1_{i+1}}^n; Y_{2_i} | W_1, Y_2^{i-1})$$
$$+ I(W_2; Y_{2_i} | W_1, Y_{1_{i+1}}^n, Y_2^{i-1}) + I(Y_{2_i}; W_2 | W_1, Y_2^{i-1}) + n\varepsilon \quad (A9)$$

$$= \sum_{i=1}^{n} I(Y_{1_i}; W_1, Y_2^{i-1} | Y_{1_{i+1}}^n) - I(Y_{1_{i+1}}^n; Y_{2_i} | W_1, W_2, Y_2^{i-1})$$
$$+ I(W_2; Y_{2_i} | W_1, Y_{1_{i+1}}^n, Y_2^{i-1}) + n\varepsilon \quad (A10)$$

$$\leq \sum_{i=1}^{n} I(Y_{1_i}; W_1, Y_{1_{i+1}}^n, Y_2^{i-1}) + I(W_2; Y_{2_i} | W_1, Y_{1_{i+1}}^n, Y_2^{i-1}) + n\varepsilon \quad (A11)$$

$$= \sum_{i=1}^{n} I(W_2, X_{2_i}; Y_{2_i} | W_1, X_{1_i}, Y_{1_{i+1}}^n, Y_2^{i-1}, X_{3_i})$$
$$+ I(W_1, Y_{1_{i+1}}^n, Y_2^{i-1}, X_{3_i}, X_{1_i}; Y_{1_i}) + n\varepsilon \quad (A12)$$

$$= \sum_{i=1}^{n} I(X_{2_i}; Y_{2_i} | X_{1_i}, V_{12_i}, X_{3_i})$$
$$+ I(X_{1_i}, V_{12_i}, X_{3_i}; Y_{1_i}) + n\varepsilon \quad (A13)$$

where (A7) follows from [10], (A8), (A9) and (A10) follow from the chain rule, (A11) follows since mutual information is non-negative, (A12) follows from the fact that $X_{1_i}$ is a function of $W_1$, $X_{2_i}$ is a function of $(W_1, W_2)$, and $X_{3_i}$ is a function of $Y_2^{i-1}$, and (A13) follows from the Markov chain $(Y_{1_i}, Y_{2_i}) - (X_{1_i}, X_{2_i}, X_{3_i}) - (W_1, W_2)$ and $V_{12_i} \triangleq (Y_{1_{i+1}}^n, Y_2^{i-1}, W_1)$.

At last, the upper bounds (A5) and (A13) can be transformed into the single-letter bounds according to the standard steps (see [5]). The bounds (IV.1.A), (IV.1.C) and (IV.1.D) are proved in [19].

## REFERENCES

[1] A. Schrijver, *Theory of Linear and Integer Programming*. New York: Wiley, 1998.
[2] T. M. Cover, and J. Thomas, *Elements of Information Theory*. Wiley-Interscience, 2006.
[3] G. Kramer, *Topics in Multi-User Information Theory*, ser. Foundations and Trends in Communications and Information Theory. Vol. 4: No. 45, pp 265-444, 2008.
[4] K. Marton, "A coding theorem for the discrete memoryless broadcast channel," *IEEE Trans. Inf. Theory*, vol. IT-25, no. 3, pp. 306–311, May 1979.
[5] T. M. Cover, and A. El Gamal, "Capacity theorems for the relay channel," *IEEE Trans. Inf. Theory*, vol. IT-25, pp. 572–584, May 1979.
[6] S. Gel'fand, and M. Pinsker, "Coding for channels with random parameters," *Probl. Contr. and Inf. Theory*, vol. 9, no. 1, pp. 19–31, 1980.
[7] T. S. Han, and K. Kobayashi, "A new achievable rate region for the interference channel," *IEEE Trans. Inf. Theory*, vol. IT-27, pp. 49–60, Jan. 1981.
[8] H. F. Chong, M. Motani, and H. K. Garg, "Backward decoding strategies for the relay channel," *MSRI Workshop: Mathematics of Relaying and Cooperation in Communication Networks*, Apr. 2006.
[9] N. Devroye, P. Mitran, and V. Tarokh, "Achievable Rates in Cognitive Radio Channels," *IEEE Trans. Inf. Theory*, vol. 52, pp. 1813–1827, May 2006.
[10] Y. Liang and G. Kramer, "Rate regions for relay broadcast channels," *IEEE Trans. Inf. Theory*, vol. 53, no. 10, pp. 3517–3535, Oct. 2007.
[11] I. Maric, R. Dabora, and A. Goldsmith, "On the capacity of the interference channel with a relay," in *Proc. IEEE International Symposium on Information Theory (ISIT 2008), Toronto, Canada*, pp. 554–558, Jul. 2008.
[12] J. Jiang, and Y. Xin, "On the achievable rate regions for interference channels with degraded message sets," *IEEE Trans. Inf. Theory*, vol. 54, pp. 4707–4712, Oct. 2008.
[13] A. Goldsmith, S. Jafar, I. Maric, and S. Srinivasa, "Breaking spectrum gridlock with cognitive radios: An information theoretic perspective," in *Proc. IEEE*, 2009.
[14] S. I. Bross, "On the discrete memoryless partially cooperative relay broadcast channel and the broadcast channel with cooperating decoders" *IEEE Trans. Inf. Theory*, vol. 55, pp. 2161-2182, May. 2009.
[15] I.-H. Wang, and D. N. C. Tse, "Interference mitigation through limited receiver cooperation," *submitted to IEEE Trans. Inf. Theory*, Nov. 2009.
[16] S. Rini, D. Tuninetti, and N. Devroye, "State of the cognitive interference channel: a new unified inner bound, and capacity to within 1.87 bits," *2010 International Zurich Seminar on Communications*, Mar. 2010.
[17] H.-Y. Chu, "On the achievable rate regions for a class of cognitive radio channels: interference channel with degraded message sets with unidirectional destination cooperation," in *Proc. IEEE International Symposium on Information Theory (ISIT 2010), Austin, Texas, USA*, pp. 445–449, Jun. 2010.
[18] V. Prabhakaran and P. Viswanath, "Interference Channels with Destination Cooperation," *IEEE Trans. Inf. Theory*, vol. 57, pp. 187-209, Jan. 2011.
[19] H.-Y. Chu, and H.-J. Su, "On the Cognitive Interference Channel with Unidirectional Destination Cooperation," *preprint,* Feb. 2011. Available: http://arxiv.org/abs/1102.3127